\documentclass[11pt]{article}
\pdfoutput=1 

\usepackage{jcappub} 

\usepackage[T1]{fontenc} 
\usepackage{multirow}
\usepackage{epstopdf}

\title{Ultra faint dwarf galaxies: an arena for testing dark matter versus modified gravity}
\author{Weikang Lin}
\author{and Mustapha Ishak}
\affiliation{Department of Physics, University of Texas at Dallas, Richardson, TX 75083, USA}
\emailAdd{wxl123830@utdallas.edu}
\emailAdd{mishak@utdallas.edu}

\abstract{The scenario consistent with a wealth of observations for the missing mass problem is that of weakly interacting dark matter particles. However, arguments or proposals for a Newtonian or relativistic modified gravity scenario continue to be made. A distinguishing characteristic between the two scenarios is that dark matter particles can produce a gravitational effect, in principle, without the need of baryons while this is not the case for the modified gravity scenario where such an effect must be correlated with the amount of baryonic matter. We consider here ultra-faint dwarf (UFD) galaxies as a promising arena to test the two scenarios based on the above assertion. We compare the correlation of the luminosity with the velocity dispersion between samples of UFD and non-UFD galaxies, finding a significant loss of correlation for UFD galaxies. For example, we find for 28 non-UFD galaxies a strong correlation coefficient of $-0.688$ which drops to $-0.077$ for the $23$ UFD galaxies. Incoming and future data will determine whether the observed stochasticity for UFD galaxies is physical or due to systematics in the data. Such a loss of correlation (if it is to persist) is possible and consistent with the dark matter scenario for UFD galaxies but would constitute a new challenge for the modified gravity scenario.}

\begin{document}
\maketitle
\flushbottom


\section{Introduction}
Very early, observations of clusters of galaxies and galaxies seemed to infer a dynamical mass much larger than the observed luminous mass. The evidence from observations for such a mismatch only continued to grow over the decades leading to the well established problem of dark matter, see for example a partial list of reviews \cite{1997review-Primack,Rev:GalacticDM,Rev:AstroParticles,Rev:DarkMatterParticle}.

The presence of weakly interacting dark matter particles has been so far a consistent scenario to explain various observations. These include
the galactic velocity dispersions in clusters \cite{zwicky1933spectral,smith1936Virgomass},
the flat rotation curves in spiral galaxies \cite{1994review-Galacitic-dark-matter},
the gravitational lensing observations \cite{2013CHFTlens,2015planck-lensing},
the cosmic microwave background anisotropy profile \cite{2015planck-params},
the small fraction of the baryonic matters inferred from primordial deuterium abundance after big bang nuclear synthesis \cite{2003BBN},
the bottom-up cosmological structure formation scenario \cite{2015review-cosmological-structure-formation},
the lensing and X-ray images separation in the bullet cluster \cite{bulletcluster-NASA},
and (assuming at least some of the dark matter are WIMP particles) the excess of $\gamma$-ray radiation from the Galactic center and Reticulum II \cite{2014-GalacticExcess,2015-RetII-gamma-ray,2015-RetII-gamma-rayII}. There is also a weak signal of $\gamma$-ray excess from the direction of Tucana III which may have a dark matter annihilation origin \cite{2016DM-annihilation-TucanaIII}.

Problems associated with the dark matter scenario such as the missing satellites and the `too-big-to-fail' problem have been attributed in the literature with the need of a better understanding of the dynamics of the dark matter and baryonic matter at dwarf galaxy scales \cite{2015review-cosmological-structure-formation,missing-satellites1,missing-satellites2,toobigtofail,Sim2015MNRAS-TBTFproblems}. Similarly, problems with density profile cusps at the center of spiral galaxies have been associated with limitations in numerical simulations \cite{2010CuspsCore}.

On the other hand, it was put forward that the missing mass problem could be an indication of modified gravity\footnote{Here we use modified gravity to refer to gravity theories proposed toward the dark matter problem.} \cite{Rev:2012Mond,Mond1983Milgrom,MondandTeVeS2004,moffat2006scalar}. In such a scenario, either the gravitational force or the law of dynamics is modified so that, for example, the rotation curves appear to be flat at large radii. Although strong constraints and criticism have been put over the years against modified gravity scenarios for the missing mass problem, renewed arguments and proposals continue to appear in the scientific literature \cite{Rev:2012Mond}. In addition, recently the mimetic gravity approach that was first studied in \cite{2013mimetic-origin} is able to explain the flat rotation curves \cite{2015mimetic-rotation-curves}.

In order to test the dark matter particle scenario, a plethora of experiments have been designed and operating over the years in order to detect such particles if they are to weakly interact with baryonic matter \cite{Rev:DarkMatterParticle,Rev:LHCdarkmatter,direct-search}. Although good progress has been made in narrowing the dark matter parameter space, no detection have been made so far. Furthermore, it is possible that, besides gravitational interaction, dark matter particles are self-interacting and that has been also probed indirectly by studying any resulting radiation \cite{Rev:AstroParticles,2014-GalacticExcess,2015-RetII-gamma-ray,2015-RetII-gamma-rayII}.

Another approach to the question is to examine the dark matter as a source of gravity which can, in principle, be segregated from baryonic matter, leading to gravitational effect without baryonic matter. This is in contrast with the modified gravity scenario which require a consistent presence of baryonic matter in order to exhibit gravitational phenomena. The bullet cluster result belongs to this type \cite{bulletcluster-NASA}, in which two clusters collided and the dark matter are ahead of the hot gas, leading to a separation of the lensing and X-ray image. The bullet cluster is then thought to be a strong evidence for the existence of the dark matter. However, the velocities of the colliding clusters have been questioned in the bullet cluster system as being too fast by Ref. \cite{2010BulletClusterProblem,2008Simulating-BulletCluster}.

In this paper, we propose a test of the second type above. We show that the ultra-faint dwarf (UFD) galaxies can provide a stringent test and evidence for dark matter instead of a modification to gravity. This could be in part motivated by the peculiar chemical and star formation history of UFD galaxies \cite{2014formationUFDs1,2014UFDsQuenching}. For examples, according to \cite{missing-satellites2,2014formationUFDs1,2014UFDsQuenching} the star formation in the `fossil' dwarf galaxies are `quenched' by reionization via suppressing the gas accretion or even `boiling out' the gas, making those `fossil' dwarf galaxies ultra-faint. Such a small scale could be the new laboratory for testing our understanding of dark matter and models of stars formation \cite{CommonMassDwarfs}. Using the currently available data, we find that the gravitational dynamics in UFD galaxies appear to be independent of the amount of baryonic matter compared to the non-ultra-faint dwarf galaxies, which would be consistent with rather the dark matter scenario.

\section{Analysis}
In the dark matter scenario, it has been shown in \cite{CommonMassDwarfs,LeastLuminousSegueI} from a maximum likelihood analysis that the dwarf galaxies they analyzed roughly share a common dynamical mass within $0.3~kpc$ of $M_{0.3kpc}\sim10^7M_{\odot}$ despite the fact that their luminosity spans five orders of magnitude. Similarly, according to \cite{2010method-HalfLightMass}, most of the dwarf galaxies might have the same total mass of $3\times10^9M_{\odot}$ and the less luminous dwarf galaxies do not necessarily have less total masses. Also, it was stressed in \cite{2015SimulationI,2015SimulationII,2014formationUFDs1} that baryon physics is important for halo and galaxy formation at the scale of dwarf galaxies and below. For example, the high-resolution simulation with consideration of the reionization suppression and supernova feedback show that the method of abundance matching needs to be corrected for low-mass haloes ($\sim3\times10^9M_{\odot}$) because only a fraction of low-mass haloes host galaxies \cite{2015SimulationI}, and the fraction decreases with the halo mass. From their Fig. 4, we can see that the stellar masses are more scattered for the low-mass haloes. In the simulations of \cite{2015SimulationII}, the very low-mass haloes show scattered stellar amounts, but as they state, it is not sure if such a stochasticity is physical or due to the poor resolution in the simulations of those small systems. Not all simulations reveal stochastic baryonic amount in dwarf galaxies. For example, the simulation in \cite{2016-sim-Oman-etal} found that the baryonic mass correlates strongly with the maximum circular velocity for dwarf galaxies, although the resolved systems are not as small as UFDs. Nonetheless, from the above, it is at least possible in the dark matter scenario that a loss of correlation between the stellar mass (which can be inferred from the luminosity) and the total dynamical mass for the small mass halos can occur for the low-mass haloes and we explore this here using currently available UFD galaxy data.


On the other hand, modified gravity scenario will have a clear prediction where the luminosity is well-correlated to the total mass. Indeed, in this scenario, the amount of baryonic matter is responsible for both the total mass and the luminosity. Moreover, UFD galaxies are poor in gas as we discussed above \cite{Rev:GalacticDM,2014formationUFDs1} which makes the link between the luminosity and the baryonic mass stronger.

\renewcommand{\arraystretch}{1.15}
\begin{table}[tbp]
\centering
\begin{tabular}{|c|l|l|l|c|}
\hline
names  & $~~~~~~~~~~~M_V$ & $~~~~~\sigma_v~(km/s)$ & $~~~~~r_h~(pc)$ & ~~~~~references\\ \hline
Segue &	$~~~-1.5~\pm~0.8$	& $~~~~~3.9~^{ +0.8}_{	-0.8}$	& $~~~~~29^{+8}_{-8}$ &\multirow{16}{*}{\cite{2012data-full-dwarfs-and-UFDs}}\\  \cline{1-4}
Ursa Major II&	$~~~-4.2~\pm~	0.6$	 & $~~~~~6.7~^{+	1.4}_{	-1.4}$	&$~~~~~149^{+	21}_{-	21}$ & \\ \cline{1-4}
Bootes II	&$~~~-2.7~\pm~	0.9$	&$~~~~~10.5~^{	+7.4}_{	-7.4}$	&$~~~~~51^{+	17}_{	-17}$& \\ \cline{1-4}
Segue II&	$~~~-2.5~\pm~	0.3$ &	$~~~~~3.4~^{	+2.5}_{	-2.1}$	&$~~~~~35^{+	3}_{	-3}$& \\ \cline{1-4}
Willman I&	$~~~-2.7~\pm~	0.8$	 & $~~~~~4.3~^{+	2.3}_{	-1.3 }$& 	$~~~~~25^{+	6}_{	-6}$& \\ \cline{1-4}
Coma Berenices&	$~~~-4.1~\pm~	0.5$&	$~~~~~4.6~^{+	0.8}_{	-0.8}$&	$~~~~~77^{+	10}_{-	10}$& \\ \cline{1-4}
Bootes III&	$~~~-5.8~\pm~	0.5$&	$~~~~~14~^{	+3.2}_{	-3}$   & $~~~~~$-----& \\ \cline{1-4}
Bootes	&$~~~-6.3~\pm~	0.2$ &	$~~~~~2.4~^{	+0.9}_{	-0.5}$&	$~~~~~242^{+	21}_{	-21}$& \\ \cline{1-4}
Ursa Major&	$~~~-5.5~\pm~	0.3$&	$~~~~~7.6~^{+	1}_{	-1}$&	$~~~~~319^{+	50}_{-	50}$& \\ \cline{1-4}
Herules&	$~~~-6.6~\pm~	0.4	$& $~~~~~3.7~^{+	0.9}_{	-0.9}$&	$~~~~~330^{+	75}_{	-75}$& \\ \cline{1-4}
Leo IV&	$~~~-5.8~\pm~	0.4$&	$~~~~~3.3~^{	+1.7}_{	-1.7}$&	$~~~~~206^{+	37}_{	-37}$& \\ \cline{1-4}
Canes Venatici II&	$~~~-4.9~\pm~	0.5	$& $~~~~~4.6~^{+	1}_{	-1}$&	$~~~~~74^{+	14}_{-	14}$& \\ \cline{1-4}
Leo V&	$~~~-5.2~\pm~	0.4$&	$~~~~~3.7~^{+	2.3}_{	-1.4}$&$~~~~~	135^{+	32}_{-	32}$& \\ \cline{1-4}
Andromeda XII&	$~~~-6.4~\pm~	1.2$&	$~~~~~2.6~^{+	5.1}_{	-2.6}$&	$~~~~~304^{+	66}_{-	66}$& \\ \cline{1-4}
Andromeda X&	$~~~-7.6~\pm~	1$&	 $~~~~~3.9~^{	+1.2}_{	-1.2}$&	$~~~~~265^{+	33}_{-	33}$& \\ \cline{1-4}
Andromeda XIII&	$~~~-6.7~\pm~	1.3$&	$~~~~~9.7~^{	+8.9}_{	-4.5}$&	$~~~~~207^{+	23}_{-	44}$& \\ \hline
Pisces II&	$~~~-5.0~\pm~	0.5$&	$~~~~~5.4~^{+	3.6}_{	-2.4}$&	$~~~~~58^{+	7}_{	-7}$& \cite{2015disc-spectroscopic-of-HydraII-and-PiscesII, 2012data-full-dwarfs-and-UFDs}\\ \hline
Reticulum II&	$~~~-3.6~\pm~	0.1$&	$~~~~~3.3~^{+	0.7}_{	-0.7}$&	$~~~~~55^{+	5}_{	-5}$& \cite{2015disc-spectroscopy-of-ReticulumII}\\ \hline
Horologium I&	$~~~-3.4~\pm~	0.1$&	$~~~~~4.9~^{+	2.8}_{	-0.9}$&	$~~~~~30^{+	4.4}_{-	3.3}$& \cite{2015RetIIandHorologiumISpec} \\ \hline
Triagullum II&	$~~~-1.8~\pm~	0.5$&	$~~~~~5.1~^{	+4}_{	-1.4}$&	$~~~~~34^{+	9}_{	-8}$& \cite{2015disc-spec-TriII,2015disc-TriangulumII} \\ \hline
Tucana II&	$~~~-3.8~\pm~	0.1$&	$~~~~~8.6~^{	+4.4}_{	-2.7}$&	$~~~~~165^{+	28}_{	-19}$& \multirow{2}{*}{\cite{2015TucanaIIandGrusIspec}} \\ \cline{1-4}
Grus I&	$~~~-3.4~\pm~	0.3$&	$~~~~~2.9~^{	+6.9}_{	-2.1}$&	$~~~~~62^{+	30}_{	-14}$& \\ \hline
Draco II&	$~~~-2.9~\pm~	0.8$&	$~~~~~2.9~^{	+2.1}_{	-2.1}$&	$~~~~~19^{+	8}_{	-6}$& \cite{2015DracoIIspec,2015disc-three-pan-starrs}\\ \hline
\end{tabular}
\caption{\label{UFD-data-table}Table of properties and references of the ultra-faint dwarf galaxies used in this work.}
\end{table}
\renewcommand{\arraystretch}{1.0}

This leads us to consider what quantity can be a reliable and model-independent tracer of the total mass. We first recall some of the basic observed (intrinsic) quantities of dwarf galaxies consisting of the luminosity $L$ or the (absolute) magnitude $M$ (calculated from distance modulus and apparent magnitude) , the (2D projected) half-light radius $r_h$, the ellipticity $\epsilon$, the luminosity weighted stellar velocity dispersion $\sigma_v$ and the metallicity $[F/H]$ \cite{2012data-full-dwarfs-and-UFDs}. Some other quantities can be constructed from these basic quantities. For example, in Newtonian gravity, it has been shown that the dynamical mass within the 3D unprojected half-light radius $r_{1/2}$ can be simply calculated from $\sigma_v$ and $r_h$ by $M_{1/2}(\leq r_{1/2})=930\times \frac{r_h}{1pc}\times\frac{\sigma_v^2}{km^2/s^2}$ \cite{2010method-HalfLightMass}. This $M_{1/2}$ is widely used since it doesn't depend on the dispersion anisotropy for a wide range of models. And it has been shown in Ref. \cite{2012data-full-dwarfs-and-UFDs} that $M_{1/2}$ for all dwarf galaxies demonstrate a good power law relation with their luminosity $L$. However, $M_{1/2}$ doesn't tell us the total mass since we don't know the extent of the dark matter halo \cite{Rev:GalacticDM}.
Also, we want to use a model independent quantity to be the total mass tracer. We choose to use the stellar velocity dispersion. In a system with only baryonic matter the stellar velocity dispersion and the total mass should be positively correlated whether in Newtonian gravity or in modified gravity. This logic is not strictly right if dark matter is present, since it has been shown that most dwarf galaxies might share the same total mass while their velocity dispersions are different \cite{2010method-HalfLightMass}. Even for systems only with baryonic matter, other factors like the concentration and dispersion anisotropy also affect the velocity dispersion. However, baryonic-matter-only systems with larger total masses should be more likely to have larger velocity dispersion. In other words in baryonic-matter-only systems, total masses and the velocity dispersions should be positively correlated, no matter in which gravity model.

It follows that in the dark matter scenario, it is not unexpected to see for UFD galaxies a possible loss of correlation between the stellar velocity dispersion and the luminosity (or magnitude). However, in the modified gravity scenario the stellar velocity dispersion must be correlated with the luminosity (or magnitude).

\begin{figure}[!tbp]
\centering
\includegraphics[width=0.99\textwidth]{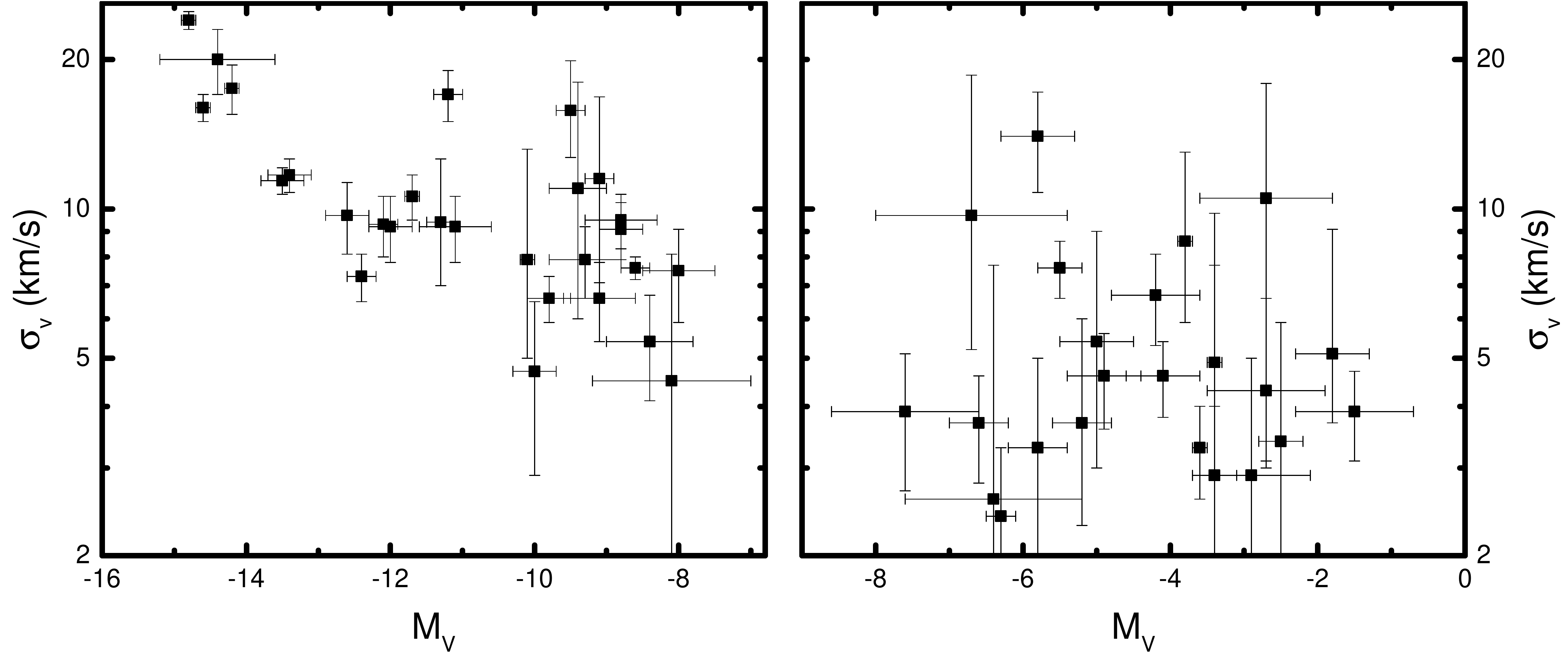}
\caption{\label{Mv-sigmav} Stellar velocity dispersion (in logarithmic scale) vs. Magnitude for non-UFD galaxies on the left and UFD galaxies on the right. LEFT: A negative correlation is found for the 28 available non-ultra-faint dwarf galaxies with a correlation coefficient of $-0.684$ (for $\log_{10}(\sigma_v)$). RIGHT: The correlation between the stellar velocity dispersion and magnitude is lost for the $23$ available UFD galaxies with a correlation coefficient dropping to  $-0.015$ (for $\log_{10}(\sigma_v)$). The results are summarized in table \ref{Mv-sigma-cc} and table \ref{table-Mv-sigmav-compare}.}
\end{figure}
\section{Results}
We analyzed the stellar velocity dispersion versus the magnitude for $28$ (non-ultra-faint) dwarf galaxies as summarized in  \cite{2012data-full-dwarfs-and-UFDs} (see references therein) and compared the results to those of $23$ UFD galaxies as available in current published data \cite{2012data-full-dwarfs-and-UFDs,2015disc-spectroscopic-of-HydraII-and-PiscesII,2015disc-spectroscopy-of-ReticulumII,2015disc-spec-TriII,2015DracoIIspec,
2015TucanaIIandGrusIspec,2015RetIIandHorologiumISpec}. In order to search for the loss of correlation mentioned above (if it exists), it is important to filter out the bright dwarf galaxies since such a loss of correlation should be present only at small enough systems.
Our selection rule for UFD galaxies is $M_V>-8$ which is roughly the definition used in the literature.  We use $-16<M_V<-8$ for the non-UFD galaxies so the magnitude range is of the same width as the one for the available UFD galaxies. Among our $23$ selected UFD galaxies are Segue I, Ursa Major II, Bootes II, Segue II, Willman I, Coma Berenices, Bootes III, Bootes, Ursa Major, Herules, Leo IV, Canes Venatici II, Leo V, Andromeda X, Andromeda XII and Andromeda XIII from \cite{2012data-full-dwarfs-and-UFDs}, Pisces II from \cite{2015disc-spectroscopic-of-HydraII-and-PiscesII}, Reticulum II from \cite{2015disc-spectroscopy-of-ReticulumII}, Horologium I from \cite{2015RetIIandHorologiumISpec}, Triangulum II from \cite{2015disc-spec-TriII}, Draco II from \cite{2015DracoIIspec}, and Tucana II and Grus I from \cite{2015TucanaIIandGrusIspec}. The properties of those UFD galaxies are summarized in table \ref{UFD-data-table}. Thus we included all UFDs for which spectroscopy is available, but the luminosity cut $M_V>-8$ filters out the two brightest UFDs, namely, Canes Venatici I and Leo T. This is reasonable, because we expect the loss of correlation to occur only for faint and small galaxy systems where stochastic processes like the supernova explosion and the reionization quenching become important.

The results of our analysis using the stellar velocity dispersion are summarized in Fig.\ref{Mv-sigmav}, table \ref{Mv-sigma-cc} and table \ref{table-Mv-sigmav-compare}.
In Fig.\ref{Mv-sigmav}, we plot the stellar velocity dispersion versus magnitude for the non-ultra-faint galaxies on the left and UFD galaxies on the right. We can see a negative correlation for the non-ultra-faint galaxies, which is expected for larger galaxies in both Newtonian dynamics and modified gravity. But on the right of the figure, we find that such correlation is lost for the $23$ UFD galaxies. This is quantitatively confirmed in table \ref{table-Mv-sigmav-compare} in terms of the correlation coefficients between the magnitude $M_V$ and the stellar velocity dispersion $\sigma_v$ or its logarithm.

We recall here that the magnitude and luminosity in V-band are related by $M_V=-2.5\log_{10}(L/L_\odot)+4.83$. So a linear relation between $M_V$ and $\sigma_v$ means $\sigma_v\varpropto\log_{10}(L)$ up to an additional constant, while a linear relation between $M_V$ and $\log_{10}(\sigma_v)$ means a power law $\sigma_v\varpropto L^p$ with a constant $p$.
To quantify the correlations, we use the standard definition of the correlation coefficient $\rho_{corr}=\frac{\sigma_{xy}}{\sigma_x \sigma_y}$ of quantities $x$ and $y$, where $\sigma_x$ and $\sigma_y$ are the standard deviations of $x$ and $y$, respectively, and $\sigma_{xy}$ is their covariance.

\renewcommand{\arraystretch}{1.15}
\begin{table}[!tp]
\centering
\begin{tabular}{ |c|c|c|c| }
\hline
\hline
\multirow{2}{*}{$M_V(>)$} & \multirow{2}{*}{$N_{gal.}$} & \multicolumn{2}{c|}{Correlation coefficient}\\
\cline{3-4}
& & $M_V$ vs. $\sigma_v$ & $M_V$ vs. $\log_{10}(\sigma_v)$\\
\hline
$-16$ & $51$ & $-0.70$ & $-0.70$ \\
\hline
$-14$ & $47$ & $-0.57$ & $-0.60$ \\
\hline
$-12$ & $41$ & $-0.53$ & $-0.55$ \\
\hline
$-10$ & $35$ & $-0.46$ & $-0.48$ \\
\hline
$-9$ & $29$ & $-0.27$ & $-0.28$ \\
\hline
\hline
$-8$ & $23$ & $-0.077$ & $-0.015$ \\
\hline
$-7$ & $22$ & $-0.13$ & $-0.055$\\
\hline
$-6$ & $18$ & $-0.29$ & $-0.27$ \\
\hline
$-5$ & $13$ & $-0.10$ & $-0.14$\\
\hline
$-4$ & $10$ & $-0.12$ & $-0.069$\\
\hline
\end{tabular}
\caption{\label{Mv-sigma-cc} In order to find the magnitude location where a significant drop of correlation occurs, we compute the correlation coefficients for the $M_V$ vs. $\sigma_v$ and $M_V$ vs. $\log_{10}(\sigma_v)$ for all $51$ dwarf galaxies with available spectroscopy (including non-UFD and UFD galaxies) with decreasing luminosity cuts. The correlation coefficients for lower luminosity cuts are smaller compared to the ones with higher luminosity cuts. A significant loss of correlation shows up when we have a magnitude cut $\sim-8$, which is also the rough cut for UFD definition in the literature \cite{2014UFDsQuenching,Maximum-Likelihood}. After this point, all correlation coefficients, such as those of the last five rows, are very small. We therefore find that the loss of correlation roughly happens at $M_V\sim-8$. We comment in the text on the temporary rise of the correlation coefficient at $M_{V,cut}=-6$. The contrast between the correlation coefficient for the non-UFDs and UFDs is shown in table \ref{table-Mv-sigmav-compare} further below.}
\end{table}

\begin{table}[!htbp]
\centering
\begin{tabular}{ |c|c|c|c| }
\hline
\hline
\multirow{2}{*}{Category} & \multirow{2}{*}{$N_{gal.}$} & \multicolumn{2}{c|}{Correlation coefficient}\\
\cline{3-4}
& & $M_V$ vs. $\sigma_v$ & $M_V$ vs. $\log_{10}(\sigma_v)$\\
\hline
non-UFD galaxies & $28$ & $-0.69$ & $-0.68$\\
\hline
UFD galaxies & $23$ & $-0.077$ & $-0.015$\\
\hline
\end{tabular}
\caption{\label{table-Mv-sigmav-compare}Comparison of the correlation coefficients for the $28$ non-UFD galaxies and the $23$ UFD galaxies. The coefficient drops from $-0.69$/$-0.68$ for the non-UFD galaxies to $-0.077$/$0.015$ for the UFD galaxies. This loss of correlation between $M_V$ and $\sigma_v$ appears to indicate uncorrelated amounts of the baryonic matter and the total mass of the UFD galaxies. This loss of correlation (if it persists in incoming and future data) is consistent and not unexpected within the dark matter scenario in UFD galaxies but would be inconsistent with the modified gravity scenario.}
\end{table}

In order to analyze the dependence of the correlation coefficients on the luminosity cuts, we gradually lower the cuts in table \ref{Mv-sigma-cc} and see how the coefficients change. We find that lower luminosity cuts give smaller correlation coefficients. In particular, at the $M_V\sim-8$ there is a significant drop (loss) in the correlation. This is also the rough cut used in the literature for UFDs \cite{2014UFDsQuenching,Maximum-Likelihood}. Next, comparing the correlation coefficients for the $28$ non-UFD galaxies and the $23$ UFD galaxies in table \ref{table-Mv-sigmav-compare}, we find that it drops from $-0.69$ and $-0.68$ for the non-UFD galaxies (for the linear and logarithmic proportionality respectively) to $-0.077$ and $-0.015$ for the UFD galaxies. This loss of correlation between the luminosity $L_\odot$ (related to the baryonic matter) and $\sigma_v$ (related to the total dynamical mass) in the UFD galaxies seems to indicate stochasticity in the amounts of baryonic matter within the UFD galaxies compared to their total masses, which is not consistent with modified gravity scenarios but consistent with the dark matter scenario. Finally, the correlation coefficients rises temporarily at $M_{V,cut}=-6$, but this is because such a magnitude cut happens to take out some UFD galaxies at the lower-left corner and makes the data somehow less scattered. However, we can see that the correlation coefficients drop again.

It worth verifying that this loss of correlation is not due to the large uncertainties associated with the UFDs data. To do so, we calculate the probability distribution $P(\rho_{corr})$ using Monte-Carlo simulations. We associate each data point with an uncorrelated Gaussian distribution $P_i(M_{V,i},~\sigma_{v,i})$ where $M_{V,i}$ and $\sigma_{v,i}$ are treated as parameters with mean values given by the data point and the deviation given by its uncertainty. For each realization, each parameter value is picked randomly from a range of $4\sigma$ around the mean. The total probability distribution function is then given by $P(\mathbf{M_V},\mathbf{\sigma_v})=\prod P_i(M_{V,i},~\sigma_{v,i})$. We use the Monte-Carlo Markov-Chain software \textit{COSMOMC} \cite{Cosmomc} as a generic sampler and treat $\rho_{corr}$ as a derived parameter to calculate $P(\rho_{corr})$. The resulting mean value and $1\sigma$ uncertainty for $\rho_{corr}$ are $-0.05\pm0.17$ (or $-0.22 \le \rho_{corr} \le 0.12$). This result is  consistent with the $-0.077$ calculated above directly from the data without taking into account the uncertainties. The correlation coefficient is thus again found to be small (and consistent with $0$), indicating that the loss of correlation is likely not due to the large uncertainties associated with the data.

\begin{figure}[ht]
\centering
\includegraphics[width=0.6\textwidth]{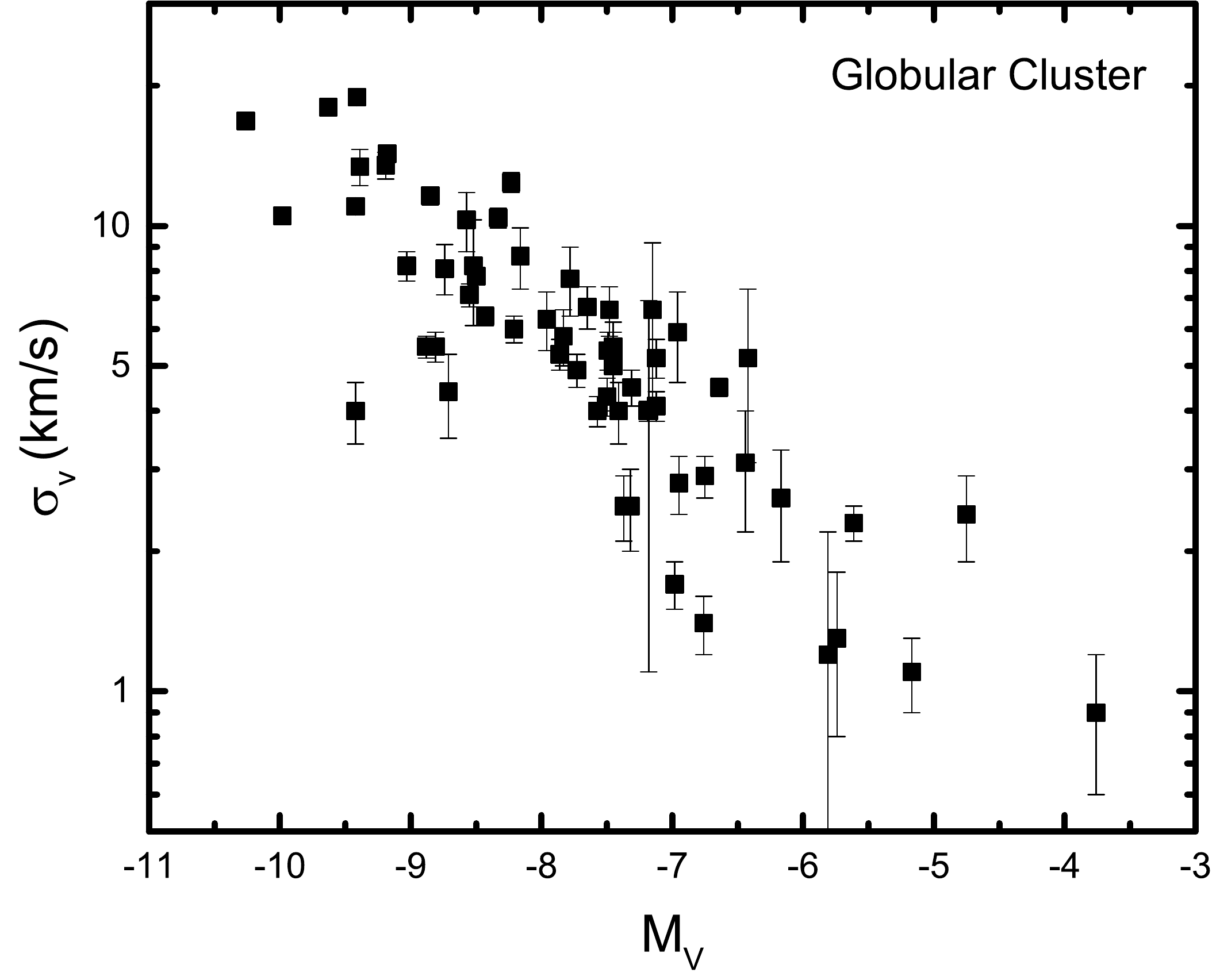}
\caption{\label{globular-cluster} The stellar velocity dispersion (in logarithmic scale) vs. the magnitude for $62$ globular clusters obtained from \cite{GloCluster}.}
\end{figure}

It is worth contrasting our result for dwarf galaxies with those of globular clusters which are considered to be dominated by stars, so their luminosity and stellar velocity dispersion should be correlated no matter how low the luminosity cut is. we plot in Fig.\ref{globular-cluster} the stellar velocity dispersion versus the magnitude of $62$  globular clusters obtained from \cite{GloCluster}. The range of magnitude in Fig.\ref{globular-cluster} is roughly from $-11$ to $-3$. This magnitude range corresponds to that of some small non-ultra-faint dwarf galaxies and some large ultra-faint dwarf galaxies. Throughout this magnitude range, $\sigma_v$ and $M_V$ stay correlated very well for globular clusters. We can also see from table \ref{Mv-sigma-globular} that the sudden drop of correlation coefficients does not occur in the globular clusters, and instead they remain well correlated regardless of the decrease of the magnitude cut.

\begin{table}[htp]
\centering
\begin{tabular}{ |c|c|c|c| }
\hline
\hline
\multirow{2}{*}{$M_V(>)$} & \multirow{2}{*}{$N_{glob.}$} & \multicolumn{2}{c|}{Correlation coefficient}\\
\cline{3-4}
& & $M_V$ vs. $\sigma_v$ & $M_V$ vs. $\log_{10}(\sigma_v)$\\
\hline
all & $62$ & $-0.78$ & $-0.86$ \\
\hline
$-9$ & $52$ & $-0.76$ & $-0.83$ \\
\hline
$-8$ & $38$ & $-0.74$ & $-0.80$ \\
\hline
$-7.5$ & $30$ & $-0.69$ & $-0.76$ \\
\hline
$-7$ & $16$ & $-0.57$ & $-0.64$ \\
\hline
$-6.5$ & $10$ & $-0.65$ & $-0.66$ \\
\hline
\hline
\end{tabular}
\caption{\label{Mv-sigma-globular}Correlation coefficients for the $M_V$ vs. $\sigma_v$ and $M_V$ vs. $\log_{10}(\sigma_v)$ for globular clusters. We can see that the velocity dispersion and the luminosity keep correlated well regardless of the magnitude cut. This feature is clearly in contrast with the ultra-faint dwarf galaxies.}
\end{table}

In view of our results, it is worth addressing the claims of Ref. \cite{Correlated-Deviations-from-BTFR} that tidal effects are the cause of deviations from MOND expectations in the Baryonic Tully-Fisher Relation (BTFR) for low mass dwarf galaxies. Ref. \cite{Correlated-Deviations-from-BTFR} uses for MOND the relation $M_b=AV_c^4$, where $M_b$ is total baryonic mass, $V_c$ is the circular velocity and $A$ is a constant. They find that the faint end of the BTFR is observed to deviate significantly from the the prediction of MOND, as shown in their Fig. 1. It is then claimed there that those deviations are not random. To try to show that, they first define a quantity called the residual as $F_b=M_b/(AV_c^4)$. So $F_b=1$ means the observation of $M_b$ matches the prediction of MOND, but if $F_b$ is greater or smaller than $1$ then the observation deviates from MOND's prediction. This $F_b$ is found significantly different from $1$ for low-mass dwarf galaxies, but is found to be "correlated" with some other quantities that are qualitatively related to how severely the dwarf galaxies experience tidal disruption from their hosts. Those correlations are then claimed to suggest: "fainter, more elliptical, and tidally more susceptible dwarfs deviate farther from the BTFR", disfavoring stochastic processes and in favor of tidal disruption.

\begin{figure}[!htbp]
\centering
\includegraphics[width=0.6\textwidth]{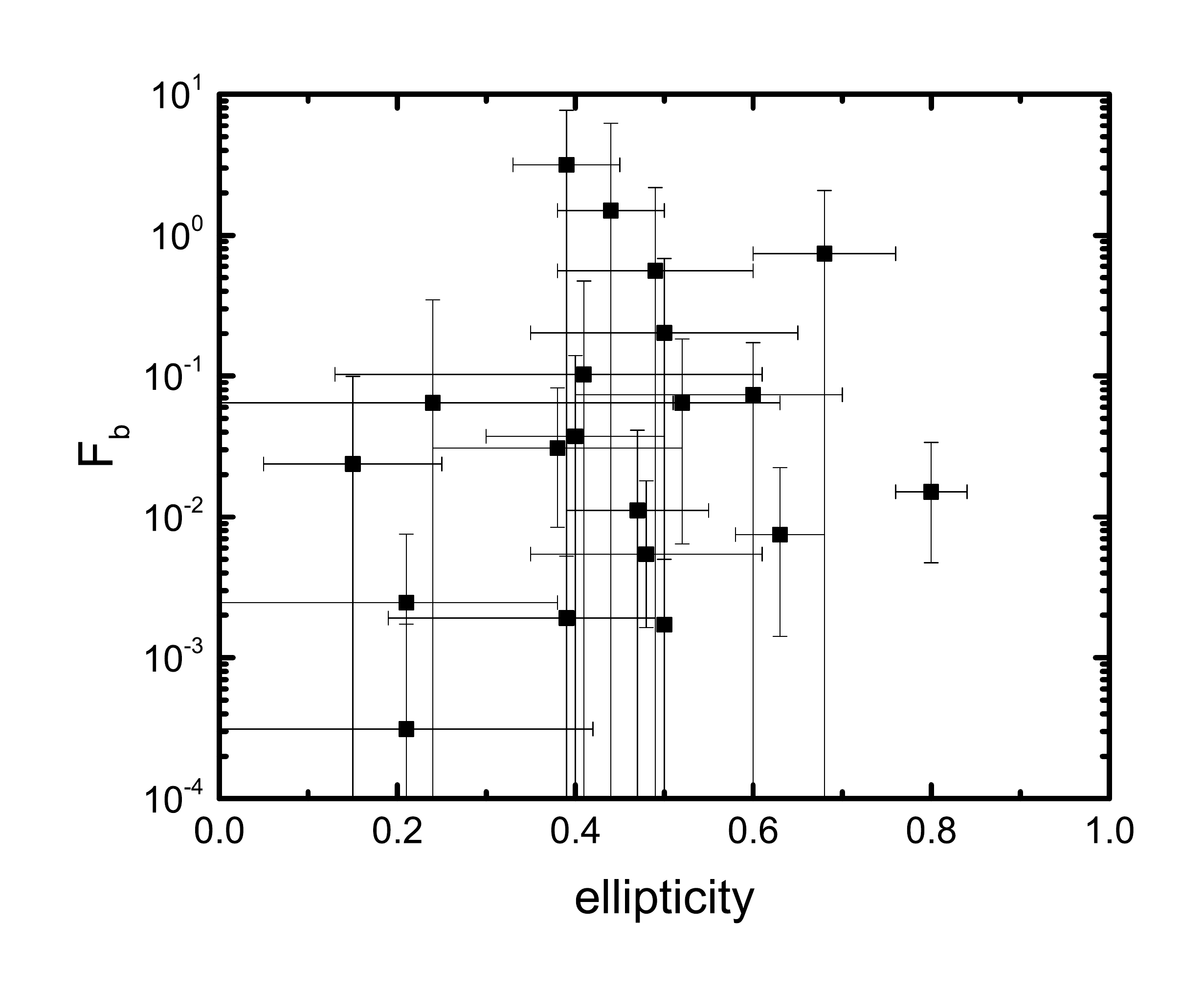}
\caption{\label{fig-Fb-e} Plot of the BTFR residual parameter $F_b$ and the ellipticity $\epsilon$. The correlation found in \cite{Correlated-Deviations-from-BTFR} is lost for the $23$ UFDs listed in table \ref{UFD-data-table}.}
\end{figure}

We don't agree with the analysis and conclusion in Ref. \cite{Correlated-Deviations-from-BTFR} for two reasons as follows. First, some quantities they used are automatically correlated with $F_b$ by construction, and consequently the corresponding correlations can't be used as evidences for the presence of tidal effects (or evidences for anything) as we explain. Indeed, their definition of $F_b$ is directly related to $M_b$. For gas poor dwarf galaxies $M_b=\Upsilon_*^VL$ where $\Upsilon_*^V$ is the mass-to-light ratio, so $\ln F_b=\ln L+\ln(\tfrac{\Upsilon_*^V}{AV^4})$ is by construction correlated with $\ln L$. The regression coefficient (the slope) for the $\ln F_b$ vs. $\ln L$ scattered plot is expected from the definition, with no surprise, to be close to $1$, as shown in the upper left panel of their Fig. 2. And since the luminosity is usually correlated with the effective radius $R_e$, $F_b$ is consequently also correlated with it as well by construction (shown in the upper right panel of their Fig. 2). Also the definition of the susceptibility to tidal influences $F_{T,D}\equiv\tfrac{M}{m}\big(\tfrac{r}{D}\big)^3=\tfrac{r^3}{GAV^2D^2}\tfrac{1}{F_b}$ is inversely related to $F_b$. It is then not surprising that $\ln F_b$ and $\ln F_{T,D}$ are anti-correlated with each other, and the slope is expected to be close to $-1$ (shown in the lower right panel of their Fig. 3). These explain some of the correlations found in \cite{Correlated-Deviations-from-BTFR} without involving tidal effects, such as $\ln F_b$ with $\ln L$, $\ln R_e$ and $\ln F_{T,D}$. Second, the elipticity and $F_b$ are not expected to be correlated with each other, but they are found to be correlated in Ref. \cite{Correlated-Deviations-from-BTFR}. However this correlation shown in Ref. \cite{Correlated-Deviations-from-BTFR} is not very strong in comparison to the other ones, and in fact does not hold for the $23$ UFDs as shown in our Fig. \ref{fig-Fb-e}. To calculate $F_b$, we however use a constant stellar mass to light ratio $\Upsilon_*^V=1.5$, but the change of correlation caused by the uncertainty of $\Upsilon_*^V$ is minimized in a logarithmic scale $\ln F_b$. Since the conclusion in \cite{Correlated-Deviations-from-BTFR} is mainly based on the above correlations, we argue that their analysis and interpretation are incorrect.

\section{Conclusion}
The results found here indicate a loss of the correlation between magnitude/luminosity of UFD galaxies and their stellar velocity dispersion compared to the non-UFD galaxies. We verified that the results are not due to large uncertainties using Monte-Carlo simulations. Incoming and future UFD galaxy data and higher resolution simulations will be able to determine whether this is physical or simply due to systematics in the data and limitations in the resolution of simulations. However, if these results persist with more data, this will be a new challenge for the modified gravity scenario to address the dark matter question and will constitute a new test for this problem. In this work we only have $23$ available UFD galaxies so one should monitor how these results will evolve as more future UFD galaxy data becomes available and explore if any systematics in the data or bias can alter the interpretation of these results. One needs to be cautious about any bias associated with structural and kinematic properties obtained from observations. For example, the low surface brightness makes UFD galaxies hard to be distinguished from the foreground and gives incorrect estimation of the structural properties, but the consistent results yielded by different groups could be comforting for now \cite{DarkMatterInDwarfs}. The bias on the velocity dispersion can be caused by the rotation, binary stars and external tidal field, all of which can `inflate' the velocity dispersion \cite{DarkMatterInDwarfs}. However, only two of our selected UFD galaxies (Herules and Ursa Major II) are potentially suffering tidal disruption \cite{2012data-full-dwarfs-and-UFDs,missing-satellites1}, and most of the dwarfs systems are pressure supported rather than rotation \cite{DarkMatterInDwarfs}. The bias from binary stars can be significant for UFD galaxies since it is of the same order of magnitude with UFD galaxies' velocity dispersion. One more possible concern is that whether this loss of correlation is caused by observational selection bias. For example, as pointed out in \cite{2010-Bullock-etal-Stealth-galaxies}, the inferred common mass $M_{0.3kpc}$ of dwarf galaxies is possibly due to the fact halos with smaller $M_{0.3kpc}$ have surface brightness too faint to be detected. However, such a selection bias is unlikely to be responsible for the loss of correlation found here as we explain. Indeed, if the loss of correlation is due to such a selection bias, one would require that on the right panel of Fig. \ref{Mv-sigmav} some luminous UFDs with larger velocity dispersions have evaded detection. This is not likely, because as shown in the right panel of Fig. 2 in Ref. \cite{2010-Bullock-etal-Stealth-galaxies}, more luminous galaxies have larger velocity dispersion detection limit. So if some luminous UFDs with larger velocity dispersion truly exist, they should have been detected and present in the top-left of the right panel of our Fig. \ref{Mv-sigmav}. Finally, it is worth noting that about a dozen of UFD candidates are found but are still waiting for spectroscopically confirmed and analyzed \cite{2015disc-eight-newDFDs,2015disc-nine-UFDs,2015disc-Pegasus,2015disc-three-pan-starrs,2015disc-horologiumII}, and one is even found outside the local group \cite{2014disc-Intracluster-UFD}. So, fortunately for the test performed here, more UFD galaxies are expected to be found in ongoing and future surveys and will allow us to investigate further this finding.

\acknowledgments
We thank
M. Kesden,
L. King,
L. Strigari,
M. Troxel,
and M. Walker for useful comments on the manuscript.
MI acknowledges that this material is based upon work supported in part by NSF under grant AST-1517768 and an award from the John Templeton Foundation.

\bibliographystyle{JHEP}
\bibliography{DMandUFDG}

\end{document}